\documentclass[preprint,5p,times,twocolumn,10pt]{elsarticle}

\usepackage{amssymb}

\journal{Physics Letters A}

\begin{document}

\newcommand{\beq}{\begin{equation}}
\newcommand{\eeq}{\end{equation}}
\newcommand{\beqa}{\begin{eqnarray}}
\newcommand{\eeqa}{\end{eqnarray}}
\def\ket#1{|\,#1\,\rangle}
\def\bra#1{\langle\, #1\,|}
\def\braket#1#2{\langle\, #1\,|\,#2\,\rangle}
\def\proj#1#2{\ket{#1}\bra{#2}}
\def\expect#1{\langle\, #1\, \rangle}
\def\trialexpect#1{\expect#1_{\rm trial}}
\def\ensemblexpect#1{\expect#1_{\rm ensemble}}
\def\kpsi{\ket{\psi}}
\def\kphi{\ket{\phi}}
\def\bpsi{\bra{\psi}}
\def\bphi{\bra{\phi}}

\begin{frontmatter}

\title{Comment on ''Entanglement transformation between two-qubit mixed states by LOCC'' [Phys. Lett. A 373 (2009) 3610]}

\author{Iulia Ghiu}
\ead{iulia.ghiu@g.unibuc.ro}
\address{Centre for Advanced Quantum Physics, University of Bucharest, PO Box MG-11,
R-077125, Bucharest-M\u{a}gurele, Romania}

\begin{abstract}
  The paper [Phys. Lett. A 373 (2009) 3610] by D.-C. Li analyzes the transformation between two-qubit mixed states by local operations and classical communication. We show that the proof of the main theorem, Theorem 2.6 in [Phys. Lett. A 373 (2009) 3610] is not complete.
Therefore the generalization of Nielsen's theorem to mixed states still remains an open problem.

\end{abstract}


\end{frontmatter}

\section{Definition of transformations between mixed states by local operations and classical communication}

In this section we give the definition of transformations by LOCC. An operation $\Lambda $ on a bipartite quantum system described by the density operator $\rho $ is called {\it separable} if it has the following form \cite{Horodecki,Gheorghiu,Sperling,Bennett}:
\beq
\Lambda (\rho )=\sum_{k=1}^N\left( A_k\otimes B_k\right) \rho \left( A_k^\dagger \otimes B_k^\dagger \right).
\label{op-transf}
\eeq
The Kraus operators have to satisfy the condition
\beq
\sum_{k=1}^NA_k^\dagger A_k\otimes B_k^\dagger B_k=I_A\otimes I_B.
\eeq

{\bf Definition.} {\it Local operations and classical communication} (LOCC) are a subset of separable operations, where the Kraus operators can be generated in a specific way. Namely, an operation described by $\{ A_i^{(1)}\}$ (with $\sum_iA_i^{(1)\dagger }A_i^{(1)}=I_A$) is applied by Alice. She communicates the result $i$ of a measurement to Bob, who uses this information for building his set of operations $\{B_j^{(2,i)} \}$. Then, Bob transmits the result $j$ to Alice and the scenario is repeated \cite{Horodecki,Gheorghiu,Sperling,Bennett}.

Using the above definition, the following results were obtained:

(i) transformation between two pure bipartite entangled states \cite{Nielsen}: $$T:\ket{\Psi}\stackrel{LOCC}{\longrightarrow } \ket{\Phi};$$

(ii) transformation between a pure state and an ensemble \cite{Jonathan}:
$$T:\ket{\Psi}\stackrel{LOCC}{\longrightarrow } \{ q_j,\ket{\Phi_j}\};$$

(iii) transformation between probability distributions (ensembles) of $2\times 2$ pure states \cite{Gour}:
$$T:\{ p_1,p_2;\ket{\Psi_1},\ket{\Psi_2}\} \stackrel{LOCC}{\longrightarrow } \{ q_1,q_2;\ket{\Phi_1},\ket{\Phi_2}\}. $$
Here the ensembles consist of two states.

The entanglement of a mixed state $\rho $ is defined as \cite{Jonathan}:
$$
E(\rho )=\min_{
\{p_i,\ket{\Psi_i}\}}
\sum_kp_kE_k(\ket{\Psi_k}),
$$
where $\rho=\sum_ip_i\proj{\Psi_i}{\Psi_i}$ is any realization of $\rho $. The ensemble that achieves the minimum in the above definition is called an optimal ensemble \cite{Li}.

Instead of using the definition presented above, the author presents a different definition for LOCC \cite{Li}:

\vspace{0.5cm}
''Let $ \{ p_i,\ket{\Psi_i}\} $ and  $\{ q_j,\ket{\Phi_j}\}$ be two optimal ensembles of two-qubit mixed states $\rho $ and $\sigma $, respectively. We say

{\bf Definition 2.5.} $\rho \stackrel{LOCC}{\longrightarrow } \sigma  $ if $\{p_i,\ket{\Psi_i}\} \stackrel{LOCC}{\longrightarrow }  \{q_j,\ket{\Phi_j}\}$.''
\vspace{0.3cm}

There is no paper in the scientific literature where this definition for LOCC for mixed states can be found.

\section{The incompleteness of the Proof of Theorem 2.6 in the paper \cite{Li}}
The main theorem in Ref. \cite{Li}, Theorem 2.6, gives the necessary and sufficient conditions that enable the transformations between mixed states using LOCC. In the body of the Proof of this theorem, the conditions that perform the transformation between ensembles are presented, where the ensembles are built using four $2\times 2$ states. The conditions that describe this transformation are given by Eqs. (9) $-$ (12) in \cite{Li}:
\newpage
\beqa
&&T_1:\ket{\Psi_1}\stackrel{LOCC1}{\longrightarrow } \{ q_j,\ket{\Phi_j}\} \hspace{0.2cm} is \hspace{0.15cm} described \hspace{0.15cm}  by \nonumber\\
&&x_1\ge p_{11}y_1+p_{21}y_2+p_{31}y_3+p_{41}y_4; \hspace{0.12cm} Eq.\hspace{0.1cm} (9)\hspace{0.1cm} in \hspace{0.1cm}  [1] \nonumber\\
&&\nonumber\\
&&T_2:\ket{\Psi_2}\stackrel{LOCC2}{\longrightarrow } \{ q_j,\ket{\Phi_j}\} \hspace{0.2cm} is\hspace{0.15cm} described\hspace{0.15cm}  by \nonumber\\
&&x_2\ge p_{12}y_1+p_{22}y_2+p_{32}y_3+p_{42}y_4; \hspace{0.12cm} Eq.\hspace{0.1cm} (10)\hspace{0.1cm} in \hspace{0.1cm} [1] \nonumber\\
&&\nonumber\\
&&T_3:\ket{\Psi_3}\stackrel{LOCC3}{\longrightarrow } \{ q_j,\ket{\Phi_j}\} \hspace{0.2cm} is\hspace{0.15cm} described \hspace{0.15cm}  by \nonumber\\
&&x_3\ge p_{13}y_1+p_{23}y_2+p_{33}y_3+p_{43}y_4; \hspace{0.12cm} Eq.\hspace{0.1cm} (11)\hspace{0.1cm} in \hspace{0.1cm}  [1] \nonumber\\
&&\nonumber\\
&&T_4:\ket{\Psi_4}\stackrel{LOCC4}{\longrightarrow } \{ q_j,\ket{\Phi_j}\} \hspace{0.2cm} is \hspace{0.15cm} described \hspace{0.15cm}  by  \nonumber\\
&&x_4\ge p_{14}y_1+p_{24}y_2+p_{34}y_3+p_{44}y_4. \hspace{0.12cm} Eq. \hspace{0.1cm} (12)\hspace{0.1cm} in \hspace{0.1cm}  [1]   \nonumber
\eeqa
As one can see from the above equations, the input pure state $\ket{\Psi_i}$ is known. There are four different LOCC transformations $T_1$, $T_2$, $T_3$, and $T_4$ that are performed, depending on the input state. One can apply the methods used in the proof of Theorem 1 in Ref. \cite{Jonathan} and the proof of the Nielsen's theorem in \cite{Nielsen} for obtaining the mathematical expression of the four transformations. If the input state is $\ket{\Psi_1}$, then one can construct a local protocol $L_1$ that makes the transformation $\ket{\Psi_1}\to \ket{\overline \eta }$, where $\ket{\overline \eta }$ is the average target state defined in Ref. \cite{Jonathan}. Further a POVM is applied to the average state in order to obtain the ensemble: $\ket{\overline \eta }\to \{ q_j,\ket{\Phi_j}\}$. Due to the fact that the local protocols $L_1$, $L_2$, $L_3$, and $L_4$ are different, depending on the input state $\ket{\Psi_i}$, it follows that the transformations $T_1$, $T_2$, $T_3$, and $T_4$ are different.

The equations (9)$-$(12) in Ref. \cite{Li} are analogue (identical) with the Eq. (9) in Ref. \cite{Gour} by Gour, where only two states were considered. Gour emphasized that there are two different transformations $T_1$ and $T_2$, that enable the LOCC between ensembles of two states.

 The author concluded in the Proof of Theorem 2.6 in \cite{Li} that, if the Eqs. (9)$-$(12) are satisfied, then $\rho \stackrel{LOCC}{\longrightarrow } \sigma $. Or, in other words, if the four different transformations $T_1$, $T_2$, $T_3$, and $T_4$ exist, then $\rho \stackrel{LOCC}{\longrightarrow } \sigma $. This statement is not so obvious and further investigation is required in order to prove it.

In the case when the transformation is applied to the mixed state $\rho=\sum_{i=1}^4p_i\proj{\Psi_i}{\Psi_i}$, the input state $\ket{\Psi_i}$ is unknown to Alice and Bob. According to the definition of LOCC presented in Sec. 1, a unique transformation $T$ described by the Kraus operators $\{A_k\otimes B_k \}$ given in Eq. (\ref{op-transf}) have to be found for the four states $\ket{\Psi_1}$, $\ket{\Psi_2}$, $\ket{\Psi_3}$, $\ket{\Psi_4}$. In addition, classical communication is used by Alice and Bob to generate the transformation. One has to prove that if Eqs. (9)$-$(12) in Ref. \cite{Li} are satisfied, then a unique LOCC transformation $T$ exists, regardless of the input state $\ket{\Psi_i} $ and this is equivalent to the transformation $\rho \stackrel{LOCC}{\longrightarrow } \sigma $.

 In conclusion, we have shown that the proof of Theorem 2.6 in Ref. \cite{Li} is not complete.
 The generalization of the Nielsen's theorem to mixed states still remains an open problem even in the simple case of two-qubit states.

\section*{Acknowledgement}
I wish to thank the two anonymous referees for their valuable suggestions, which led to an improved version of this Comment. This work was supported by CNCSIS - UEFISCSU, postdoctoral research project PD code 151/2010 for the University of Bucharest.

\end{document}